\documentclass[11pt]{article}
\usepackage{epsfig}
\usepackage{amssymb}
\newcommand{\arcmin}{$^{\prime}$}

\newcommand{\arcsec}{$^{\prime \prime}$}
\title{{\bf Dwarf Galaxies in the Leo I Group: the Group
Luminosity Function beyond the Local Group}} 
\author{K.~Flint$^1$, M.~Bolte$^1$, C. Mendes de Oliveira$^2$\\
\vspace{0.1cm}\\
\normalsize $^1$UCO/Lick Observatory, University of California, Santa Cruz, CA 95064\\
\normalsize $^2$Instituto Astr\^{o}nomico e Geof\'{i}sico (IAG), Av. Miguel 
Stefano 4200, CEP: 04301-904,\\ S\~{a}o Paulo, Brazil\\
}
\date{}
\begin{document}
\maketitle
\def\bull{\vrule height .9ex width .8ex depth -.1ex}
\makeatletter
\def\ps@plain{\let\@mkboth\gobbletwo
\def\@oddhead{}\def\@oddfoot{\hfil\tiny
``Dwarf Galaxies and their Environment'';
International Conference in Bad Honnef, Germany, 23-27 January 2001}%
\def\@evenhead{}\let\@evenfoot\@oddfoot}
\makeatother

%%  if your contribution is short, you may, if the title is clear enough, 
%%  skip the abstract.....
\begin{abstract}\noindent
We present first results of a survey of the Leo I group at 10 Mpc for
$M_R \le -10$ 
dwarf  galaxies.  This is part of a larger program to
measure the faint end of the galaxy luminosity function in nearby poor
groups.  Our method is optimized to find Local-Group-like dwarfs down
to dwarf spheroidal surface brighnesses, but
we also find very large LSB dwarfs in Leo I with no Local Group
counterpart. 
A preliminary measurement of the luminosity function yields a slope
consistent with that measured in the Local Group.
\end{abstract}

\section{Introduction}

The classical picture of the dependence of the luminosity function
(LF) on environment is that richer environments have steeper faint-end
slopes. 
The range of measured slopes, using $\alpha$ of the Schechter (1976)
formalism, was considered to vary from $\alpha \simeq -1.4$ in rich
clusters to 
$\alpha \simeq -1$ for the low-density field.  Galaxy groups have long
been included with the field in these models, taking the Local Group
(LG) slope of $\alpha = -1.1$ (Pritchet \& van den Bergh 1999) as the
prototype.  
As the LFs are
superpositions of the LFs of individual morphological types,
this suggests that dwarf galaxies contribute a larger fraction by
luminosity in denser environments.  

  Yet, there have a been
a number of observational constraints on determining the LF to dwarf
luminosities of $M_B > -16$, where this steepening occurs.  Many
photometric surveys do not reach much fainter than this potential
turn-up at $M_B \sim -16$, as 
dwarfs are notoriously difficult to detect at any significant distance.
This has often led to a trade-off in group and cluster studies
between the distance of a galaxy sample (and thus the limiting magnitude)
and angular sky coverage, as 
the nearer groups and clusters cover substantial and almost unwieldy
fractions of the sky.  This leads to further difficulties in
membership classification, as the more distant groups and clusters
require statistical membership determination using control fields
for the faintest galaxies which can prove problematic ({\em c.f.}
Valotto, Moore, \& Lambas 2001).  Finally, as we probe fainter dwarf
galaxies, we also probe fainter surface brightnesses which introduce
significant surface-brightness selection
effects.

\section{Our Survey}
With our $R$-band survey of the Leo I group, we probe a nearby poor
group to observational limits approaching those of the LG.
Leo I is a poor group at a distance of 10 Mpc and contains NGC 3379
as a member.  The best work on the LF of the group to date is the
photographic work of Ferguson \& Sandage (1991).  These data
 reached a limiting magnitude of $M_B = -14.2$ (adjusted for
$m - M = 30$), which for a $\langle B - R \rangle = 1.3$ corresponds
to $M_R = -15.5$.  Our program is designed to extend these limits.

The strategy of our survey has three main features which mitigate some
of the more insidious observational difficulties. 
 First, 
our imaging survey uses the KPNO 0.9m+MOSAIC, which has a
59\arcmin$\times$59\arcmin\ field of view with eight mosaiced CCDs.
Thus, we have
$R$-band imaging of over seven square degrees in Leo I
with the
advantage of the linear response of CCDs.  Second, the proximity
of Leo I at 10 Mpc allows for galaxy membership classification on a
galaxy-by-galaxy basis, using morphology, photometric parameters,
radial profiles, colors, and in many cases, directly measured
distances.   
Finally, we have developed a detection technique that optimizes
detection of very low-surface-brightness (LSB) dwarfs, extending our
survey to both faint luminosities and faint surface brightnesses which 
approach the limits measured in the Local Group.

Our detection method is two-fold. First we use the traditional method of
standard SExtractor detection to find  high-surface-brightness
(HSB) objects.  We then complement this with our optimized method, 
which is based on the work of Dalcanton (1995) for finding large LSB
galaxies in the field. We mask our image of
high-surface-brightness features, and then convolve it with a filter
of the shape and size we expect for dwarf galaxies at 10 Mpc.  In
Figure 1a, we show an example of a dwarf in our sample that is not
detected by our traditional method.  In the central panel, we show the
masked image, and in the right panel, the same image convolved with a
5\arcsec\ exponential kernel.  The
aperture size shown is the detection aperture from the optimized method,
where the dwarf is now a significant detection.  Our method is 
described more fully in Flint et al. (2001).

\begin{figure}[ht]
\begin{center}
\includegraphics[width=4in]{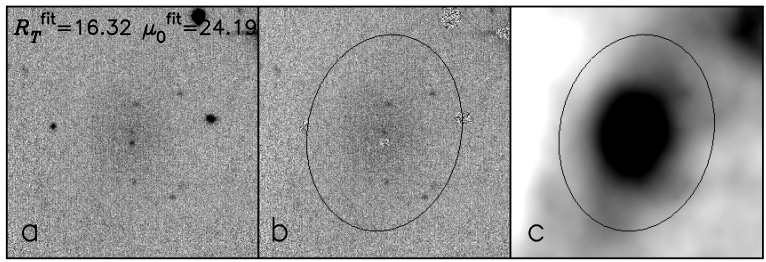}
\end{center}
\caption{{\bf a:} Dwarf in Leo I, undetected by traditional
method. {\bf b:} The image masked of HSB features. {\bf c:} The
image convolved with 5\arcsec\ exponential kernel, now showing up as a
significant detection.}
\end{figure}

\section{Selection Function}
An advantage of our detection procedure is that it is completely
automated; thus, we can run extensive Montecarlo simulations to
tune our detection parameters, quantify our completeness, and
calculate our errors in measuring photometric 
parameters.  We generate artificial galaxies,
input them to our data images, and apply  both our
detection methods to recover them.   We then calculate a recovery
fraction as a function of both input central surface brightness ($\mu_0$) and
input total magnitude ($R_T$).  Our simulations for one field are shown in
Figure 2, where the greyscale indicates recovery fractions of 90\%
(darkest), 70\%, 50\%, 30\%, and 10\% (lightest).  
LG galaxies,
if seen at a distance of 10 Mpc, are plotted for comparison.
While we don't detect dwarfs like
Draco and Ursa Minor, we find that without our optimized method, we
would only detect objects with $\mu_0 < 23.5$ and so would miss dwarfs
like And II and fainter.  Furthermore, these data were taken with the
MOSAIC's engineering grade chips which are difficult to flat field.
We estimate that with flatter data, we could extend our method one
magnitude in total magnitude and two magnitudes in central surface
brightness.   Yet, even with these limitations, we find that at the 90\%
completeness level we can find dwarfs similar to Antlia and Sculptor,
and at the 50\% completeness level, dwarfs similar to Tucana and Leo
II (Flint et al. 2001).

\begin{figure*}[ht]
\centering
\begin{minipage}[c]{0.4\textwidth}
 \centering
 \caption{Selection Function for one field. Greyscale indicates
recovery fractions of 90\%
(darkest), 70\%, 50\%, 30\%, and 10\% (lightest).  The dotted lines
are lines are lines of constant exponential scale length, and the
solid lines are lines of constant isophotal size at a limiting surface
brightness of 26.7 $R$ mag/$\prime \prime$.}
\end{minipage}%
\hspace{0.2cm}
\begin{minipage}[c]{0.45\textwidth}
 \centering
 \includegraphics[width=0.8\textwidth]{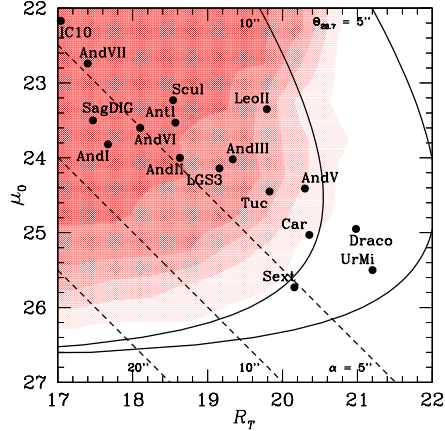}
\end{minipage}
\end{figure*}

\begin{figure*}[hb]
\centering
\begin{minipage}[c]{0.4\textwidth}
 \centering
 \caption{Group candidates for 80\% of the imaging data. Crosses are
 objects detected with the traditional method, while stars are objects
 detected only through the optimized method. Dotted and solid lines
 are the same as in the previous figure, and the filled dots are
 comparison Local Group galaxies.}
\end{minipage}%
\hspace{0.2cm}
\begin{minipage}[c]{0.45\textwidth}
 \centering
 \includegraphics[width=\textwidth]{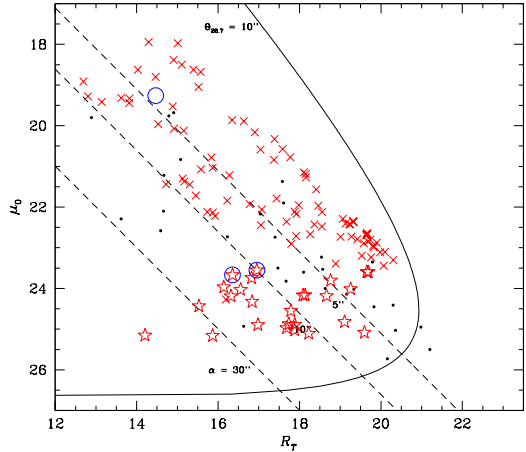}
\end{minipage}
\end{figure*}

\section{First Results}
First results from Leo I are shown in Figure 3.  Here we plot our
detections in the same way as Figure 2, where the filled dots are
LG galaxies again for comparison.  Here the crosses are
objects detected via our traditional method.  The stars are objects
which would not have been detected without using the optimized
method. These detections have had a preliminary membership
classification, using their position in this figure, profile type,
and morphology.  Using our sample of follow-up observations, we find
that objects falling in the upper right area of the figure typically
are small, higher-redshift background objects.  We find some degree of
contamination from background spirals around the Freeman's Law value
of $\mu_0 
\simeq 20$, where the open circle indicates a background spiral removed
from the sample.  
Ironically, however, we find the most robust
membership classification so far for the lowest-surface-brightness
objects, where the circled stars are examples of dwarfs we have
identified as members via spectroscopic redshifts and
surface-brightness fluctuations (SBF).  
Our follow-up program is on-going and
includes velocity measurements from both HI and optical spectroscopy,
SBF, and colors for all candidates.  In this way, we also hope to
guard against 
contamination from field LSB galaxies, cosmologically
dimmed high-$z$ 
galaxies, and possibly diffuse light from $z \gtrsim 0.6$ galaxy
clusters.

An interesting feature of our sample is that we find a few galaxies which
deviate from the typical $R_T -
\mu_0$ relation followed by the LG galaxies in Figure 3.
These galaxies, if members, are large, LSB dwarfs not seen in the
LG.  Similar galaxies have previously been discovered in other
environments such as Virgo (Impey, Bothun, \& Malin 1988)
and M81 (Caldwell et al. 1998). 

\section{Luminosity Function}
With the data in Figure 3, we can begin to construct the group
LF.  In Figure 4, in the shaded histogram, we show
the  raw galaxy counts for the 80\% of the data analyzed to date.
We then weight these raw counts for incompleteness
as quantified through the Montecarlo simulations, as a function of
both $\mu_0$ and $R_T$.  The scaled LF is shown as the open histogram
in Figure 4.  A preliminary measurement of the faint-end slope
yields 
$\alpha = -1.2$, which is consistent with $\alpha = -1.1$ measured in
the Local Group  (Pritchet \& van den Bergh 1999).  

\begin{figure}
\centering
\begin{minipage}[c]{0.4\textwidth}
 \centering
 \caption{Preliminary luminosity function for 80\% of Leo I data,
 adopting $m-M=30$. Filled histogram: raw galaxy counts. Open
 histogram: galaxy counts scaled for completeness as a function
 of ($\mu_0, R_T$) using results from the simulations.}
\end{minipage}%
\hspace{0.2cm}
\begin{minipage}[c]{0.45\textwidth}
 \centering
 \includegraphics[width=0.8\textwidth]{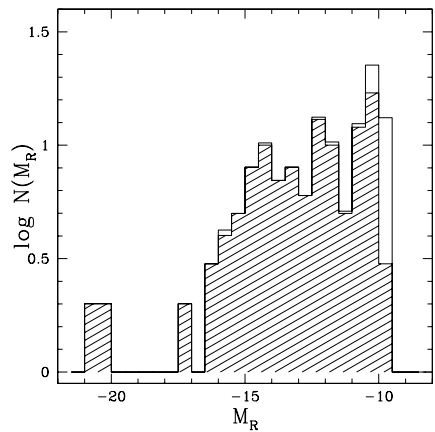}
\end{minipage}
\end{figure}

\section{Summary}
We present a new program for robustly detecting
low-luminosity dwarfs
at a distance of 10 Mpc in the Leo I group.  Using an
optimized, filter-detection technique for finding
low-surface-brightness dwarfs, we probe the group luminosity function
to $M_R \simeq -10,\ \mu_0=24.5$ at the 50\% completeness level.  We
use follow-up observations and morphological membership classification
to construct a preliminary luminosity function which appears to be
consistent with that of the Local Group.  We also find several large,
LSB dwarfs which, if they are members, deviate from the $R_T - \mu_0$
relation and have no counterpart in the Local Group.

% References:
%%  Use the ApJ, AJ, new A\&A style (are the same!)
%%
{\small
\begin{description}{} \itemsep=0pt \parsep=0pt \parskip=0pt \labelsep=0pt
\item {\bf References}

\item
Caldwell, N., Armandroff, T. E., Da Costa, G. S., \& Seitzer, P. 1998,
AJ, 115, 535 
\item
Dalcanton, J. J. 1995, PhD Thesis, Princeton University 
\item
Ferguson, H. C., \& Sandage, A. 1991, AJ, 101, 765
\item
Flint, K., Metevier, A.J., Bolte, M., \& Mendes de Oliveira,
C. 2001, ApJS, in press (astro-ph/0101276) 
\item
Impey, C., Bothun, G., \& Malin, D. 1988, ApJ, 330, 634
\item
Pritchet, C. J., \& van den Bergh, S. 1999, AJ, 118, 883
\item
Schechter, P. 1976, ApJ, 203, 297
\item
Valotto, C. A., Moore, B., \& Lambas, D. G. 2001, ApJ, 546, 157

\end{description}
}

\end{document}